\documentclass[11pt]{article}
\usepackage{moriond,epsfig}

\def\be{\begin{equation}}
\def\ee{\end{equation}}
\def\bea{\begin{eqnarray}}
\def\eea{\end{eqnarray}}

\begin{document}
\begin{flushright}
UPRF-2002-07
\end{flushright}
\vspace*{4cm}
\title{PERTURBATIVE AND NON-PERTURBATIVE ISSUES \\IN HEAVY QUARK
FRAGMENTATION~\footnote{Talk given at the 37th Rencontres de Moriond on 
QCD and Hadronic Interactions, Les Arcs, France, 16-23 March 2002.}}

\author{ Matteo CACCIARI }

\address{Dipartimento di Fisica, Universit\`a di Parma, Italy,  and\\
INFN, Sezione di Milano, Gruppo Collegato di Parma\\[10pt]
\epsfig{file=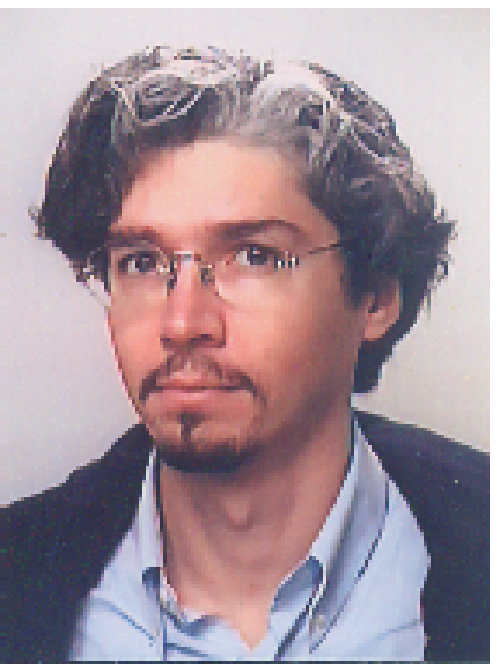,width=35mm}}

\maketitle\abstracts{
We review the state-of-the-art of our understanding of heavy
quark fragmentation. Recent $e^+e^-$  data for $B$ mesons are compared to 
the most
up-to-date theoretical predictions, and the need for inclusion of a
non-perturbative component is discussed. Experimental analyses in
moments space are suggested, and it is pointed out how
perturbative and non-perturbative contributions are to be properly
matched. Failure to do so can result in large phenomenological
discrepancies. An example is given for $B^+$ hadroproduction at the Tevatron.
}

\section{Introduction}

This talk will be devoted to a review of our present understanding of
heavy quark fragmentation, i.e. of the processes where a heavy quark is
produced in a hard collision and then 
hadronizes before decaying. For the quark to be ``heavy'', its mass 
has to be larger than the QCD scale $\Lambda_{QCD} \simeq 200-300$ MeV. 
The top quark, however, is too heavy and decays weakly 
before hadronizing. Throughout our discussion ``heavy quark'' will
therefore have to be understood as a charm or a bottom quark.

\begin{figure}
\begin{center}
\epsfig{file=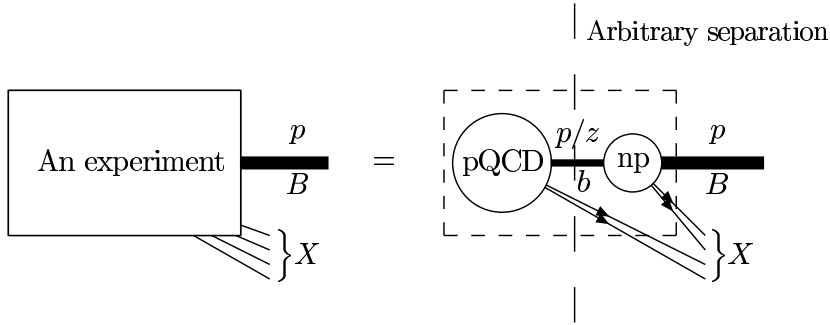,angle=90,width=11cm}
\caption{
\label{fig1} Schematic view of how the production process of an observed
$B$ meson can be separated into perturbative and non-perturbative
contributions.}
\end{center}
\end{figure}

Let us start by considering, for the sake of definiteness, an experiment
observing a $B$ meson. Since we know it to contain a $b$ quark, whatever
the initial state of the process we can envision its production to
proceed as roughly sketched in Fig.~\ref{fig1}: A so-called hard
process, described by perturbative QCD (pQCD),  produces a bottom quark.
Subsequently, soft strong interactions (the non-perturbative `np'
blob) bind this heavy quark to light ones
into a hadronic state, resulting into the observed $B$ meson with only a
fraction $z$ of the original momentum of the $b$ quark.

From such a simplified description alone, it appears clear that any
separation between ``hard'' and ``soft'' processes, which is to say
between perturbative and non-perturbative QCD, is at best arbitrary.
We shall focus more on this issue in the following.

People knowledgeable with {\sl light} quark fragmentation only might not be
immediately  familiar with the following fact: In the heavy quark case
the pQCD term  does not need collinear factorization to be performed.
It is the large mass $m$ of the heavy quark that takes care of
regulating the collinear singularities, acting as an infrared cutoff. 
Hence, pQCD alone can yield a finite result and be predictive at
the `leading twist' level. Further discrepancies with the
experimental results (aside, of course, unknown higher order
perturbative contributions) can then be  attributed to a
non-perturbative part, to be studied in the form of power corrections.

Many analyses, and also intuitive arguments, predict for the leading
term of such power corrections a $\overline{\Lambda}/m$ behaviour, 
$\overline{\Lambda}$ being
a hadronic scale of the order of a few hundred MeV. These corrections
are therefore parametrically small, $m$ being larger than
$\overline{\Lambda}$, but
can be numerically large. For instance, they are definitely larger than
typical power corrections in event shapes studies at LEP, where the
suppressing scale is $\sqrt{s} \simeq 90$~GeV rather than the heavy
quark mass. Heavy quark fragmentation is therefore an ideal place to
observe and analyze such corrections.

\section{Perturbative Issues}

Let us consider the production of the heavy quark $Q$ in $e^+e^-$
collisions, $e^+e^- \to \gamma/Z \to Q + X$. An experimentally
observable variable usually looked at is the energy fraction (with respect to 
the beam energy) of a heavy
hadron $H$ containing the heavy quark, $x_E \equiv E_H/E_{beam}$.

Fixed order perturbative 
calculations~\cite{Nason:1997nw,Rodrigo:1997gy,Bernreuther:1997jn}
are available up to order $\alpha_s^2$. 
These results do however contain
large logarithmic terms which need to be resummed to all
orders:\begin{itemize}
\item $\log(s/m^2)$ terms are large when the centre-of-mass energy
$\sqrt{s}$ is much larger than the heavy quark mass $m$. Such a
situation is easily met at LEP energies for both charm and bottom
production;
\item $1/(1-x_E)$ and $\log(1-x_E)/(1-x_E)$ terms are large when the energy of
the observed particle is close to the maximum allowed one, and are due
to gluon
radiation being inhibited close to the phase space boundaries.
\end{itemize}

Mele and Nason~\cite{Mele:1990cw} first considered the resummation 
of $\log(s/m^2)$ terms
up to next-to-leading logarithmic (NLL) level. They achieved it
 be rewriting the $e^+e^- \to \gamma/Z \to Q + X$
differential cross section in a factorized form~\footnote{From now on we
shall make use of Mellin moments, $D_N \equiv \int_0^1
x^{N-1}\;D(x)\;d(x)$, which turn convolutions into products.},
\begin{equation}
\sigma_N(\sqrt{s},m) = C_N(\sqrt{s},\mu_F) D_N^{\rm ini}(\mu_F,m) + {\cal
O}((m/\sqrt{s})^p) \; .
\end{equation}
The factorization scale $\mu_F$ separates the two functions $C$ and
$D^{\rm ini}$, and
Altarelli-Parisi evolution can be used to resum the collinear
logarithms in $D^{\rm ini}(\mu_F,m)$ by evolving from an initial scale
$\mu_{0F} \simeq m$ up to the hard scale $\mu_F \simeq \sqrt{s}$, 
so that $D_N^{\rm ini}(\mu_F,m) = E_N(\mu_F,\mu_{0F}) D_N^{\rm ini}(\mu_{0F},m)$. 

It is to be noted that all the information concerning the hard
scattering process is now contained in the coefficient function 
$C_N(\sqrt{s},\mu_F)$. The initial condition $D_N^{\rm
ini}(\mu_{0F},m)$, on the other hand, only contains information about
physics taking place around the heavy quark mass scale $m$. This process
independence of the initial condition function has been recently
established in a more formal setting~\cite{Cacciari:2001cw}. It is
worth reminding that, while process-independent, $D_N^{\rm
ini}(\mu_{0F},m)$ is of course still factorization scheme dependent, like the
coefficient function and the Altarelli-Parisi evolution factor. Both
Ref.~\cite{Mele:1990cw} and Ref.~\cite{Cacciari:2001cw} make use of the $\rm
\overline{MS}$ scheme.

After performing the resummation of the $\log(s/m^2)$ terms one still
has to take care of the potentially terms related to suppression of
radiation close to the phase space borders. Such terms show up as large
logarithms in the large-$N$ limit of the Mellin transforms. For
instance, in this limit the initial condition reads:
\begin{eqnarray}
\label{ininlon}
D_N^{{\rm ini}}(\alpha_s(\mu_0^2);\mu_0^2,\mu_{0F}^2,m^2) \!&=&\!\!
1 +  \frac{\alpha_s(\mu_0^2)}{\pi} \, C_F\; \left[ - \ln^2N + 
\left( \ln \frac{m^2}{\mu_{0F}^2} - 2\gamma_E + 1 \right) \ln N \right.  \\
\!&+&\!\!\left.  1 - \frac{\pi^2}{6} + \gamma_E - \gamma_E^2 + \left( \gamma_E 
- \frac{3}{4} \right) \ln \frac{m^2}{\mu_{0F}^2} +
{\cal O}\left(\frac{1}{N} \right) \right] + {\cal O}(\alpha_s^2) \;\;.
\nonumber
\end{eqnarray}
Resummation for these so-called Sudakov logarithms was performed at the
leading log (LL) level~\cite{Mele:1990cw} and at NLL level, but in a
process dependent way~\cite{Dokshitzer:1995ev}. In Ref.~\cite{Cacciari:2001cw},
on the other hand, NLL resummation has been revisited in a fully process
independent way, exploiting the factorization properties of $D_N^{\rm
ini}(\mu_{0F},m)$ and providing NLL resummed expressions for both the $e^+e^-$
coefficient function and the process independent initial condition. 

The resummation of all the large logarithms present in the perturbative
calculation for $D_N(\sqrt{s},m)$ allows for a prediction which is at
the same time reliable and accurate. The dependence on the unphysical
factorization scales $\mu_F$ and $\mu_{0F}$, for instance, is greatly
reduced and under control after the Sudakov logarithms are resummed with
NLL accuracy~\cite{Cacciari:2001cw}.

\section{Non-Perturbative Issues}

Despite having achieved a reliable perturbative prediction, inclusion of  
non-perturbative effects is still mandatory
for a meaningful comparison with the experimental data, for instance the
ones recently published by the ALEPH~\cite{Heister:2001jg} and 
SLD~\cite{Abe:2002iq} Collaborations. After
calculating a cross section for, say, $b$ {\sl quark} production, the one for
$B$ {\sl meson} is obtained by convoluting it with a non-perturbative
component:
\begin{equation}
D^B_N = D^b_N D^{np}_N\; .
\end{equation}

Pretty much at odds with the available literature, I
shall consider comparisons between theory and experiments in $N$-moments
space rather than with differential distributions in $x_E$
space~\cite{Giele:2002hx,Cacciari:2002pa}.

\begin{figure}
\begin{center}
\epsfig{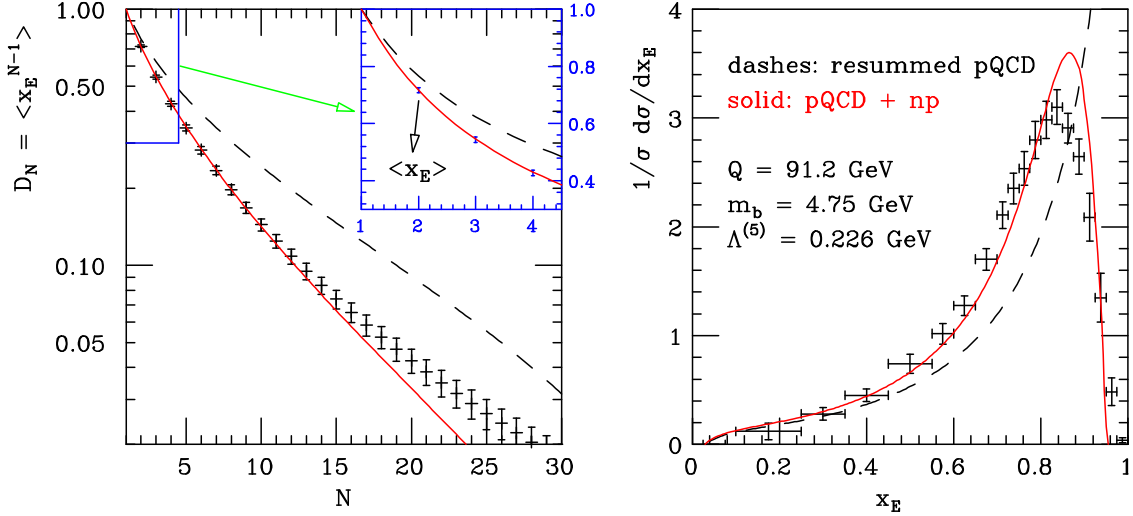}
\label{fig2}
\caption{
ALEPH~\protect\cite{Heister:2001jg} experimental 
data compared to theoretical results.
The dashed line is the perturbative prediction. The solid line is a
combination of the former with a non-perturbative component fitted to
the $D_2~=~\langle x_E\rangle$ point.}
\end{center}
\end{figure}

The most obvious reason for doing so is that only in moments space a
smooth departure of the experimental results from the {\sl perturbative}
prediction can be seen: low-$N$ moments are closest, and the gap widens
as $N$ increases. Indeed, the non-perturbative contribution can be
predicted to behave like~\cite{Jaffe:1993ie,Nason:1996pk}
\begin{equation}
D^{np}_N = 1 - (N-1)\frac{\overline{\Lambda}}{m} + {\cal
O}\left(\frac{{\overline{\Lambda}}^2}{m^2}\right)\;.
\label{leadingpc}
\end{equation}
In $x_E$ space, on the other hand, one usually tries to
compare a whole curve, and the shape of the perturbative prediction has
clearly little to do with the experimental one, especially in the $x_E
\simeq 1$ region, where the peak is.  This is due to the fact that
all-order power corrections become important here, and prompt one to
include from the very start a ``large'' non-perturbative term, for instance 
in the form of some smearing function like the Peterson et al. one, in
order to fit the data. 
Doing so, however, immediately obscures the fact that 
non-perturbative contributions can be seen as a 
small $\overline{\Lambda}/m$ correction to the perturbative result.

Figure~\ref{fig2} clearly shows this point. The purely perturbative
result fails in describing even the lowest $N$ moment (see left panel), 
hence showing
that inclusion of a non-perturbative component is mandatory. However,
its job looks much worse in $x_E$ space (right panel), particularly in the 
region around and beyond
the peak. But it suffices to fit a 1-parameter non-perturbative
form~\footnote{In this case a $D^{np}(x;\alpha) = (\alpha+1)(\alpha+2)
x^\alpha (1-x)$ form (or, rather, its Mellin transform) was used, and 
the fit returned $\alpha = 27.45$. It can easily be seen that this form
is compatible with the leading power correction~(\ref{leadingpc}) upon
replacing $\alpha$ with $2m/\overline{\Lambda}$. This equality suggests
$\overline{\Lambda} \simeq 350$~MeV, in line with our intuitive
expectations.} to
the $D_2 \equiv \langle x_E\rangle$ point in $N$ space, to produce a curve which
describes very well the low-$N$ region, and even $x_E$ data much better 
(albeit not perfectly). We wish to 
stress that {\sl no effort} has been made in this case
to achieve a particularly good description of the whole $x_E$
distribution, the emphasis being rather on showing that a ``reasonable''
functional form, fitted to a single point in moments space in terms of a
numerically small $\overline{\Lambda}/m \simeq 0.1$ correction, can already
produce a decent agreement. Good fits of the low-$N$ moments, on the
other hand, will be important for the issue we are  going to discuss in
the next Section.

\section{Phenomenology}

The theoretical machinery of fragmentation functions for heavy quark
can be used to make predictions for processes other than $e^+e^-$
production, exploiting the property of process independence of the
initial condition $D^{\rm ini}$. For instance,  bottom cross sections at
large transverse momentum $p_T$ in hadronic collisions can be
calculated, providing a resummation of  large $\log(p_T/m)$
terms~\cite{Cacciari:1993mq,Cacciari:1998it}. At
the same time, inclusion of the non-perturbative component determined
from $e^+e^-$ fits allows to make predictions for the production of $B$
mesons, which are the particles which can be directly observed. 

Obtaining accurate theoretical predictions requires however great care
in at least two instances:\begin{itemize}
\item a ``non-perturbative term'' is {\bf not} an observable quantity.
It cannot be determined in absolute terms, but only {\sl relatively
to how one defines the perturbative part and its parameters}. Therefore,
when fitting, say,  $e^+e^-$ data one extracts a non-perturbative function 
which
should then be used {\sl only} together with a perturbative description of
the {\sl same kind} (leading, next-to-leading, resummed, etc.) as the one 
it has been determined with, and with the same parameters
($\Lambda_{QCD}$, $m$, ... );
\item different processes may depend on different details of the
non-perturbative contribution. For instance, the peak in $x_E$ space is
clearly the most prominent feature in $e^+e^-$ processes, but not the most
important one when calculating $B$ meson production in $p\bar p$
collisions, where instead a good determination of the {\sl moments} around
$N\simeq 4$ is crucial~\cite{Frixione:1998ma,Nason:1999ta}.
This is due to the fact that the cross
section for production a heavy quark in hadronic collisions behaves like
$d\sigma^b/dp_T \simeq A/p_T^4$. Hence, convolution with a
non-perturbative function returns:
\begin{equation}
\frac{d\sigma^B}{dp_T} \simeq \int dz\,d\hat p_T\, D^{np}(z) 
\frac{A}{\hat p_T^n} \delta(z \hat p_T - p_T) = 
\frac{A}{p_T^n}D_n^{np} \; ,
\end{equation}
with $n\simeq 4$.
\end{itemize}

\begin{figure}
\begin{center}
\epsfig{file=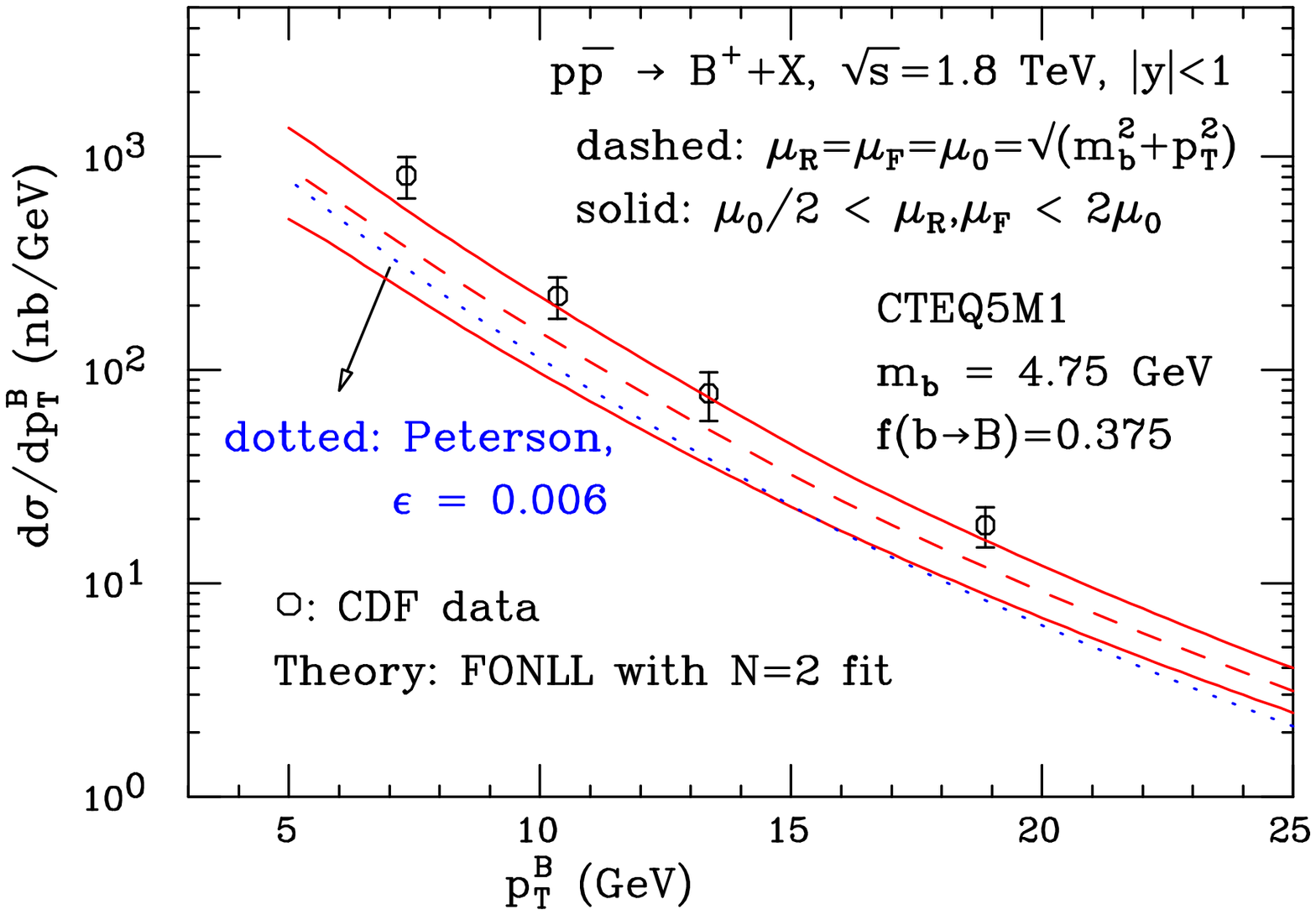,width=7.5cm,height=5.5cm}\hspace{.5cm}
\epsfig{file=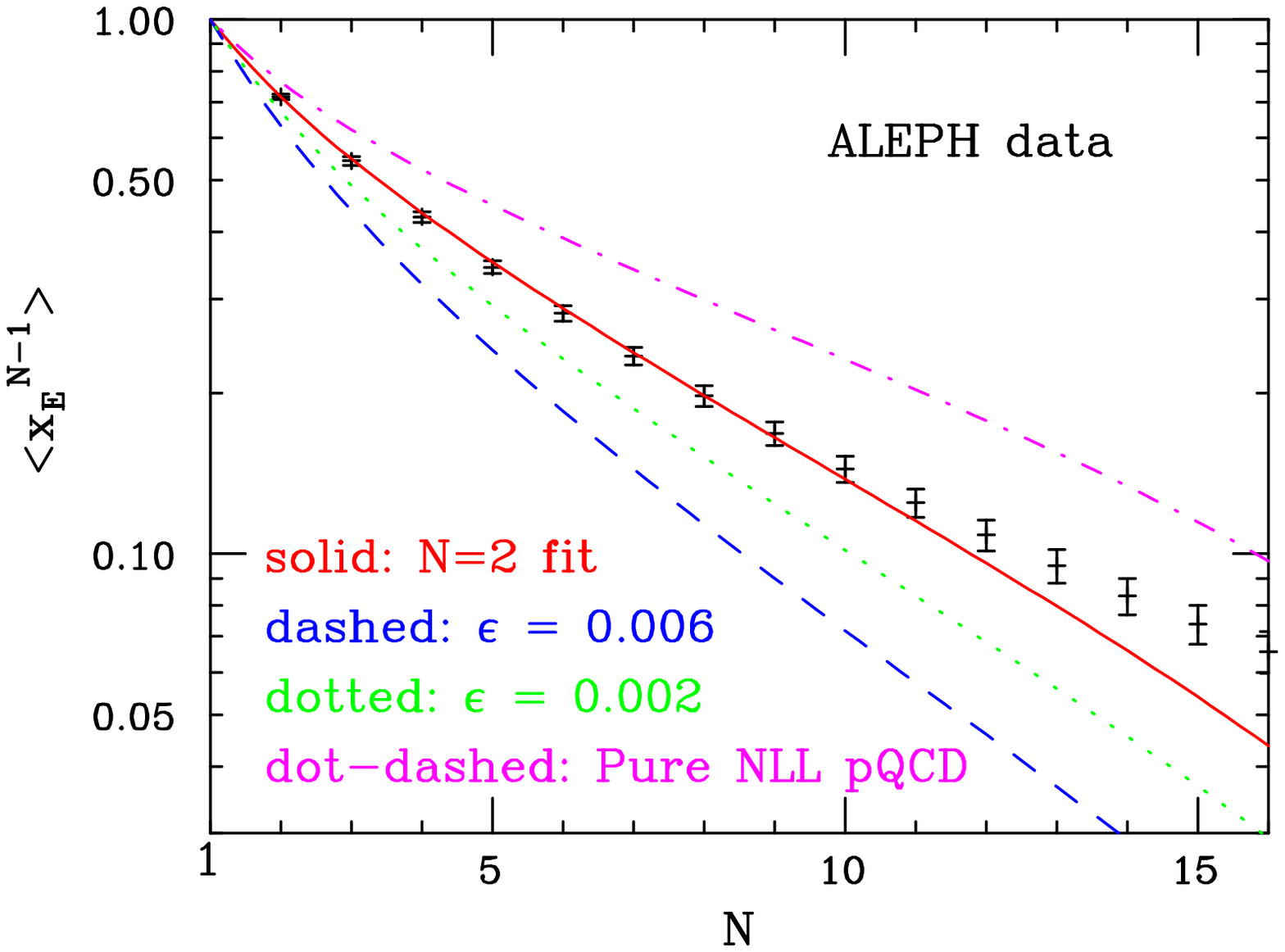,width=7.5cm,height=5.5cm} 
\caption{
\label{fig3} On the left, the experimental data from 
CDF~\protect\cite{Acosta:2001rz}
on $B^+$ production in $p\bar p$ collisions at the Tevatron, compared
to theoretical predictions. On the right, moments of 
ALEPH~\protect\cite{Heister:2001jg} data on $B$
fragmentation in $e^+e^-$ collisions compared to a purely perturbative
theoretical prediction and to various models/fits for non-perturbative
contributions. Plots from Ref.~\protect\cite{Cacciari:2002pa}. No
Sudakov resummation was included in this case.}
\end{center}
\end{figure}

That these issues are not merely of academic interest is shown by the
case of $B$ production at the Tevatron~\cite{Acosta:2001rz}. 
Figure~\ref{fig3} shows that the
``standard'' procedure, used by the experimental Collaboration, 
of evaluating the cross section by convoluting
the  perturbative calculation for $b$ quark with a Peterson et al.
function with $\epsilon = 0.006$ underestimates the data by almost a factor
of three. 

On the other hand, employing a non-perturbative contribution fitted to
$e^+e^-$ data in {\sl moments space} (the ``$N=2$ fit''), so as to get a 
good description of
moments around $N\simeq 4$ (and taking care of employing the same
kind of perturbative description in $e^+e^-$~\cite{Cacciari:2001cw} 
and $p\bar p$~\cite{Cacciari:1998it}
processes) increases the prediction, bringing
it in agreement with the data within the errors~\cite{Cacciari:2002pa}.

\section{Conclusions}

Heavy quark fragmentation is now a mature subject. Fixed order
perturbative
calculations have been performed up to order $\alpha_s^2$, collinear and
Sudakov resummations are known up to next-to-leading logarithmic accuracy.

Comparisons to experimental data however need the addition of a
non-perturbative component. The fairly large contributions expected in
this process make it a good one to study power corrections.
Performing the fits and the comparisons in {\sl moments space} helps 
disentangling the leading power correction from the higher order ones.

Factorization of terms related to the hard
scattering from terms related to the heavy quark mass scale suggests some form
of universality for such non-perturbative terms, which can therefore be
fitted in one process and used to give predictions in another.

When transporting this information care must however be taken to
properly match the {\sl perturbative and non-perturbative terms}, which
{\sl are {\bf not} independently measurable quantities}, and to
carefully consider what details of the non-perturbative function need to
be well known. Failure to do so may result in predictions with a far
larger degree of uncertainty than one might expect.

\section*{Acknowledgments}
It is a pleasure to thank Gregory Korchemsky for the invitation to give
this talk. I also wish to thank Mario Greco and Paolo Nason for the long
collaboration on this subject.

\end{document}